\documentclass{article}
\usepackage{lineno,hyperref}
\usepackage{float}
\usepackage{graphicx}
\usepackage{url}
\usepackage{amsmath}
\usepackage{color}
\usepackage{color,soul}
\usepackage[margin=1.3in]{geometry}

\title{Housing Market Forecasting using Home Showing Events}
\author{Yuanyuan Zha, Susan T. Parker, James J. Foster, Vadim Sokolov \thanks{All authors are of Graham School of Continuing Liberal and Professional Studies, University of Chicago. Corrisponding author's email: vsokolov@gmu.edu}}

\begin{document}

\maketitle

\begin{abstract}
Opportunities to support urban economic decision-making with analytical models are extensive in the real estate market. Both buyers and sellers face uncertainty in real estate transactions in about when to time a transaction and at what cost. In the current real estate market, both buyers and sellers make decisions without knowing the present and future state of the large and dynamic real estate market. Current approaches rely on analysis of historic transactions to price a property. However, as we show in this paper, the transaction data alone cannot be used to forecast demand. We develop a housing demand index based on microscopic home showings events data that can provide decision-making support for buyers and sellers on a very granular time and spatial scale.  We use statistical modeling to develop a housing market demand forecast up to twenty weeks using high-volume, high-velocity data on home showings, listing events, and historic sales data. We demonstrate our analysis using data from seven million individual records sourced from a unique, proprietary dataset that has not previously been explored in application to the real estate market. We then employ a series of predictive models to estimate current and forecast future housing demand. A housing demand index provides insight into the level of demand for a home on the market and to what extent current demand represents future expectation.  As a result, these indices provide decision-making support into important questions about when to sell or buy, or the elasticity present in the housing demand market, which impact price negotiations, price-taking and price-setting expectations. This forecast is especially valuable because it helps buyers and sellers to know on a granular and timely basis if they should engage in a home transaction or adjust their home price both in current and future states based on our forecasted housing demand index. 

\end{abstract}


\section{Introduction}

For real estate market stakeholders, selling or purchasing a home is an uncertain process. Sellers face uncertainty about how to price their home, how quickly it might sell, and how many offers they might expect from interested buyers in the market. Buyers face uncertainty about how many competitors they face in the market, how quickly they should submit an offer, and how to estimate a fair purchase price among many other factors when purchasing a home. In most cases, both buyers and sellers seek the guidance of real estate agents who advise them by primarily using domain knowledge derived from experience. An important contributing factor to uncertainty on the part of all market participants is the lack of timely, spatially-granular, predictive information available about future expectations for selling or buying a home. While robust housing market indices and indicators are commonly employed in the real estate market, most measures are not reported for several months after the economic event transpired nor are they predictive of future market activity. Existing indices and measures offer information that might be helpful in understanding past market trends, but they do not provide enough insights into the state of the real estate market in the next month, which is more valuable to real estate market decision makers. 
Predictive insight into the housing market would serve as a useful tool in the context of this uncertainty. Not only do market participants face the potential for varying outcomes in their home transaction, but inherent inefficiencies are a characteristic of the housing market. Research examining the real estate market indicates that it imperfectly adjusts to changes in market expectation as measured by price \cite{case2003there}. These market efficiencies stem from uncertainty about taxation, transaction costs, and availability of financing among other factors \cite{case1988efficiency}. Further inefficiency stems from pricesetting; sellers motivated by loss aversion set higher prices than they otherwise might \cite{goetzmann2006estimating}.  The market features fundamental inefficiencies in setting and accepting prices, but it is slow to adjust housing prices to reflect accurate levels of value \cite{genesove2001loss}. 
Existing real estate indices serve as market measures for real estate stakeholders, the most prominent of which is the S\&P/Case-Shiller Home Price Index. These indices use the logged price difference between a sale and subsequent sales of a single home to construct a home price index using linear regression across the U.S. and in large metropolitan areas. Repeat sales models face several criticisms, including skepticism about the representativeness of homes that undergo repeat sales to the entirety of the market as well as uncertainty as to which weighting mechanisms are most robust \cite{wallace1997construction}. Importantly, these indices feature lagged updates that occur months after transactions that are modeled and lack geographic precision. The lack of real-time information is especially important in relatively competitive urban housing markets, as is the geographic detail. Urban markets often contain more heterogeneous communities that feature diverse home and neighborhood features, rendering the choice between different communities within a city to be inherently different despite proximate occurrence \cite{goodman2003housing}. 
We investigate the potential of a unique home showings dataset to predict housing market demand in Chicago over a three-year period. Data on home showings is courtesy of ShowingTime, Inc., a growing Chicago-based real estate showing management company that schedules approximately half the real estate showings in the United States. ShowingTime provides services for scheduling showings for houses on the market. ShowingTime estimates that their listings comprise 80
We estimate housing market demand through analyzing the home showings and total number of properties on the market. By analyzing all the showing associated events for homes listed for sale in Chicago, we build several models for two different predictive tasks: a short term housing demand heat index to predict the portion of homes on the market that will be sold in the next two to five weeks; and a longer term housing demand index to predict portion of homes on the market that will be sold in the next twenty weeks. We experiment with several types of the models: linear regression to explain the housing demand using showings and home sold in previous weeks, ensemble models to combine linear regression, decision trees, and neural nets techniques, and time series models including LAR and LASSO. We show that these models can robustly predict housing demand with the objective of addressing a lack of knowledge or informed expectation about expected market prices and demand in the real estate market for housing. 

\subsection{Connection with Existing Work}
Short-term housing market prediction is not a well studied area in the literature. Some of the previous work include forecasting models that rely on macroscopic parameters, such as Google trends~\cite{wu2013future} or macroeconomic data~\cite{ng2008using, hua1996residential, bee1999evaluation, thomas2000prediction, gupta2010predicting}. Both parametric time series methods~\cite{gupta2010forthcoming, fan2010reliability} as well as non-parametric or semi parametric techniques were considered and shown to be successful for forecasting housing market parameters~\cite{gencay1996forecast}. A neural network approach was proposed in\cite{khalafallah2008neural}. A Bayesian techniques for short term forecasts for consumer sentiment based on Internet search queries was proposed in~\cite{scott2013bayesian}. Our approach relies on a very time series analysis techniques~\cite{shumway2013time}. Our approach is similar in spirit to previously considered auto-regressive models. However, the data set analyses in this paper allows to develop accurate forecasts at high spatial resolution, which is not possible with previously analyses data.

The rest of the paper is outlined as follows. Section~\ref{sec:background} reviews fundamentals of real estate market. Section~\ref{sec:data} describes the data set used for the analysis and provides exploratory analysis.  Section~\ref{sec:models} provides description of the statistical techniques used for developing housing market demand forests and illustrates our methodology for Chicago metropolitan area.  Finally, Section~\ref{sec:conclusion} concludes with directions for future research.


\section{Background}\label{sec:background}
Research on real estate market measures have frequently employed price to measure the inefficiencies and volatility of the real estate market. Some research has expanded on the repeat home sales models to investigate prices by employing differing methodologies or incorporating additional data about home characteristics. This is meant to address the potential bias in homes sold repeatedly, the methodology that the Case-Shiller Index uses, by adding data to differentiate between repeat sales as a result of improved properties as compared to unchanged properties \cite{case1991dynamics}. Considering the context of the overall market by including sales volume of single sales homes in the measures of home price is another approach to add additional data to generate market-level indices, as is employing autoregressive models to measure correlation at different time indices in the housing market \cite{nagaraja2011autoregressive}. 
While incorporating measures such as volume as well as additional home data have advanced the accuracy of real estate market indices, less reseach has centered on measuring demand as a function of home buyer behavior \cite{maier2009real}. Some examinations of home buyer turnover find that it is more sensitive than price measures to demand shocks though this is not universal [11, 12]. 
Part of the lesser emphasis in the literature on demand measured by home buyer behavior is driven by a limitation in data availability. Little data about demand for a house would have been available until recently. As consumers increasingly use the Internet in their home searches, storing the data that captures this demand is also becoming inexpensive. Initial use of online search engine data resulted in robust forecasts of the housing market with regard to home prices \cite{follain1995incorporating}. Real estate listing sites such as Zillow or Trulia have produced demand data by providing online mapping and valuation using their own search data. Reflecting the desire among real estate market participants for information about the market's current and future demand, approximately half of Zillow's site visits are intended to make use of the site's free algorithm to price and predict sale time of a home \cite{wu2013future}. Redfin, another online real estate listing company, made use of its data on consumer searches for housing by launching the industry's first demand index based on web traffic to home listing pages, home tours, and offers to purchase a home made through its website \cite{economist}.
Online real estate search data appears valuable in understanding consumer behavior in the real estate market, perhaps partially because of its proximity to the home buyer. An even more proximate measure to consumer activity would be a dataset of how many homes are not only searched, but physically shown to the consumer. This is because home buyers who are more serious potential purchasers make the effort to find a listing agent, examine homes, and physically view these homes in considering their potential purchase. Often, these home showings are scheduled online through a few home showings companies; ShowingTime and CSS are the dominant players in this market. If analyzed, the geographic flexibility and the demonstrated seriousness of consumer behavior in scheduling time to be shown a home is a potentially a reliable indicator of consumer demand.  In addition to being a reliable and accurate source of data for prediction applications, the data offers the potential to take real-time measures of demand through infrastructure that refreshes home showings on an ongoing daily basis, lending itself to timely prediction. Moreover, the data is available virtually nationwide and rapidly expanding, creating the potential for granular or macro examination of consumer demand, creating a valuable tool for stakeholders in the real estate market.

\section{Chicago Metropolitan Area Showings Data}\label{sec:data}
\subsection{Data Description}
The ShowingTime data employed spans January 1, 2011 to December 31, 2013 and comprises all homes, condominiums, and apartments listed for sale in Chicago that ShowingTime's scheduling system recorded. The resulting dataset contains 6 million property actions including showings, inspections, open houses, or any other appointment that ShowingTime's scheduling service recorded. However, the records included in the analysis were limited to home showings, which comprised 4.6 million records. The data is further restricted to exclude rental properties, retail, and properties selling bundled units together, which left 3.9 million residential properties.
\begin{figure}[H]
	\centering
	\includegraphics[width=1\linewidth]{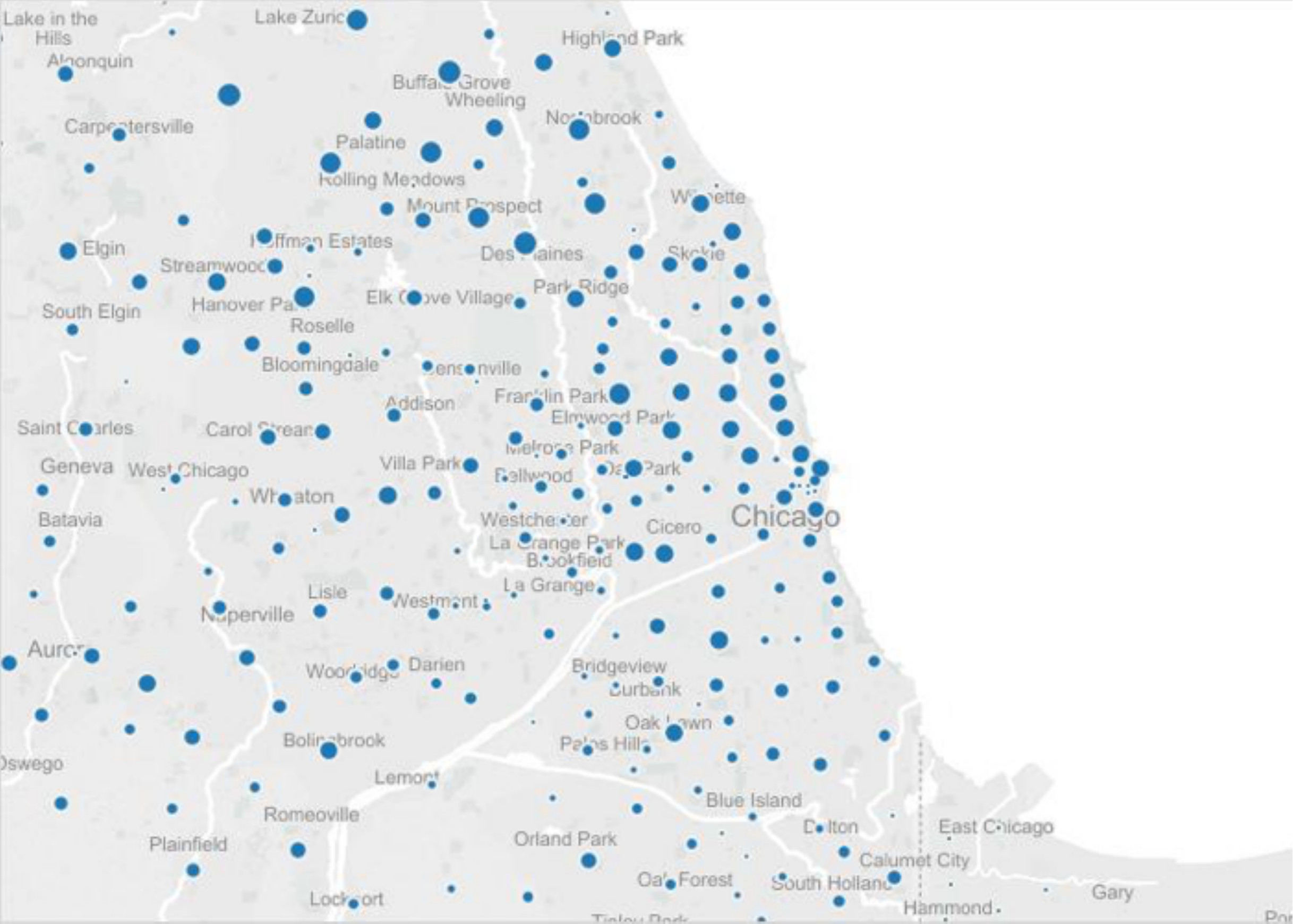}
	\caption{Geocoded Home Showings Data in the Chicagoland Area. The map above represents the volume of home showings in the ShowingTime dataset for Chicago-area home showings. Circles denote zip codes in which the homes are located and are scaled to increase with an increasing concentration of home showings. }
	\label{fig:map}
\end{figure}

The map above represents the volume of home showings in the ShowingTime dataset for Chicago-area home showings. Circles denote zip codes in which the homes are located and are scaled to increase with an increasing concentration of home showings. 

The volume of home showings for each week from 1/1/2011 to 1/1/2014 are plotted using ShowingTime data on home showings volume for the Chicagoland area. 
The home showings occurred both in the city of Chicago as well as beyond the city borders as depicted in Figure~\ref{fig:map}. The diversity in geographic areas represented in the data is both spatial and economic. Where the average home price during this time period is \$57,467 in one zip code market, the average price in another zip code market is \$330,950.

Each of the parameters listed above were aggregated to measure weekly totals that comprise the parameters of the dataset. 

To aggregate the individual showings data into variables reflecting the overall state of the housing market in Chicagoland, a series of by week indicator variables were generated. Each week of the year an event occurred corresponds to weekly dummy variables. For example, if a showing occurred during the 17th week of 2013, the home associated with that event would have a “1” in the column headed 2013 showings week 17 but not in week 16 or 18. A similar process was used to aggregate the number of homes available on the market by week, the number of homes sold by week, the median time corresponding to sold homes on market for each week, and the mean time corresponding to sold homes for each week of 2011, 2012 and 2013.
Table~\ref{tab:excerpt} is an excerpt that reflects this data structure. For each week of each year, the total number of showings, homes sold and properties listed on the market are aggregated into a single variable. For instance, for the second week of 2011, the number of showings that week numbered 11,672, the number of homes sold was 58, and the number of properties on the market in the data amounted to 15,850.
\begin{table}[H]
	\centering
	\begin{tabular}{llll}
		\hline
		Year Week & Showing & Sold & Property on Market\\\hline
		2011 week 02 & 11672 & 58 & 15850\\
		2011 week 03 & 13250 & 82 & 16153\\
		2011 week 04 & 13732 & 87 & 16410\\
		2011 week 05 & 12978 & 153 & 16637\\
		\hline
	\end{tabular}
	\caption{Excerpt From Showingtime Final Data Structure} 
	\label{tab:excerpt}
\end{table}

\subsection{Real Estate Market Demand}
Real estate demand is commonly defined as the quantity of units demanded at various prices according to a schedule of increasing demand associated with increasing prices. To understand the role of these external forces on demand for real estate, analysts often use absorption measures to detect the small shifts in demand that are characteristic of a durable goods market. Net absorption is one such index – it is the change in a real estate market's share of occupied housing stock. Average absorption is net absorption calculated over a number of years to detect meaningful trends occurring in the overall market. 
\[
Q_d = \dfrac{\Delta Q}{Q}\left/\dfrac{\Delta P}{P}\right.
\]

\begin{figure}[H]
	\centering
	\includegraphics[width=0.8\linewidth]{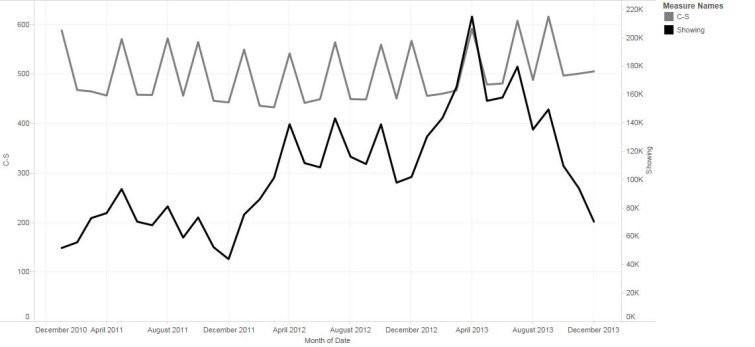}
	\caption{Time Series Plot of Showings and Case-Shiller Home Price Index. The volume of home showings for each week from 1/1/2011 to 1/1/2014 are plotted using ShowingTime data on home showings volume for the Chicagoland area as well as the Case-Shiller Home Price Index.}
	\label{fig:show-case-shiller} 
\end{figure}

In addition to the initial weekly aggregated measures, two additional variables were introduced in the context of the ShowingTime data. The first we refer to as the Housing Demand Index (HDI), which captures the number of housing sold in a week on the market as a function of total homes available during the week for sale. We model real estate demand by employing the HDI as the dependent variable in our models. 
\[
\mathrm{HDI} = \dfrac{\mbox{\# Sold}}{\mbox{\# On the market}}
\]

\begin{figure}[H]
	\centering
	\begin{tabular}{cc}
		\includegraphics[width=0.5\linewidth]{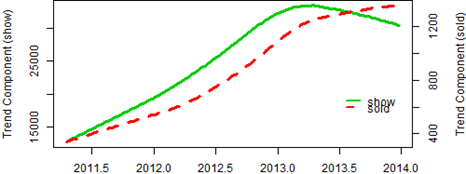} &\includegraphics[width=0.5\linewidth]{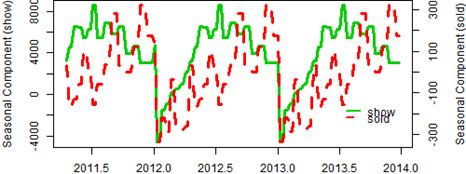}
	\end{tabular}
	\caption{Trend and Seasonal Component Comparison of Showing and Sales Data. The above plot shows the shared trend component in home showings and home sales as a function of time. The bottom plot features the seasonal component of the home showings and sales as a function of time. Showings are a solid line; sold homes are a dotted line. }
	\label{fig:trend} 
\end{figure}
A second index, the Showing Index (SI) captures the volume of showings during a week on the market as a function of the total homes available that week for sale.

\subsection{Exploratory Data Analysis}
Real estate is by nature a highly seasonal and at times cyclical process with upward or downward trending periods. The time series nature of the home showings follows the seasonal and cyclical patterns of real estate data, as shown in Figure~\ref{fig:trend}. The home showings peaks reflect the high volume summer period of home turnover and the upward trend of the data over time reflect the increased volume of home showings during the time span of the dataset. 
\begin{figure}[H]
	\centering
	\includegraphics[width=0.8\linewidth]{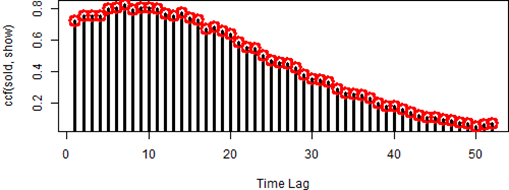}
	\caption{Cross -Correlation of Sales and Showing Data. Time lag in weeks} 
	\label{fig:cross-corr}
\end{figure}

Cross-correlation of sales data and showing data. Time lag in weeks
The untransformed HDI measure relatively closely tracks established market indices. The Case-Shiller Price Index as shown in Figure~\ref{fig:show-case-shiller} is overlaid onto the HDI measure over time . Though the HDI is a weekly measure and the Case-Shiller Price Index is a monthly measure, the two are not divergent. We observe from the relationship of the HDI to the Case Shiller Index that it features a linear relationship but further that we also observe time series seasonality. We employ linear models as well as time series models to examine how best to predict real estate market demand as modeled by our HDI measure. 
As we can see in Figure~\ref{fig:trend}, the seasonal component of both sold and showing time series are following the same trend, with highest values during the end of summer and lowest values during winter. On the other hand, the trend component of the sold time series is clearly trailing the showing time series. The seasonal component is found by taking the mean of weekly subseries (the series of all week 1, 2,… values). The seasonal values are removed, and the remainder smoothed to find the trend. The overall level is removed from the seasonal component and added to the trend component. This process is iterated a few times. 
To quantify the lagged time between the sold and showing time series, we look at cross-correlation of those two series as a function of a lag. As shown in Figure~\ref{fig:cross-corr}, the strongest relation is at lag of 10 weeks and there is a significant correlation for lags up to 20 weeks. 

\section{Analytical Models}\label{sec:models}
\subsection{Linear Models}
We present two initial linear models employing weekly aggregated home showing data: 1) a short-term heat HDI, and 2) a long-term HDI. The short-term heat HDI provides a granular prediction for the demand for homes in the Chicagoland housing market by predicting housing demand two weeks into the future, using the past ten weeks of market data. It is intended as a short-term “market heat” indicator to enable short-term decisions about demand in the real estate market. Because the inputs to the short-term heat HDI are generated from prior market data, it is possible to update the heat index in real time to generate a timely market predictor. The long-term HDI is intended to inform longer-term decision makers for whom market insight many weeks in advance is important.

\begin{table}[H]
	\centering
\includegraphics[width=0.8\linewidth]{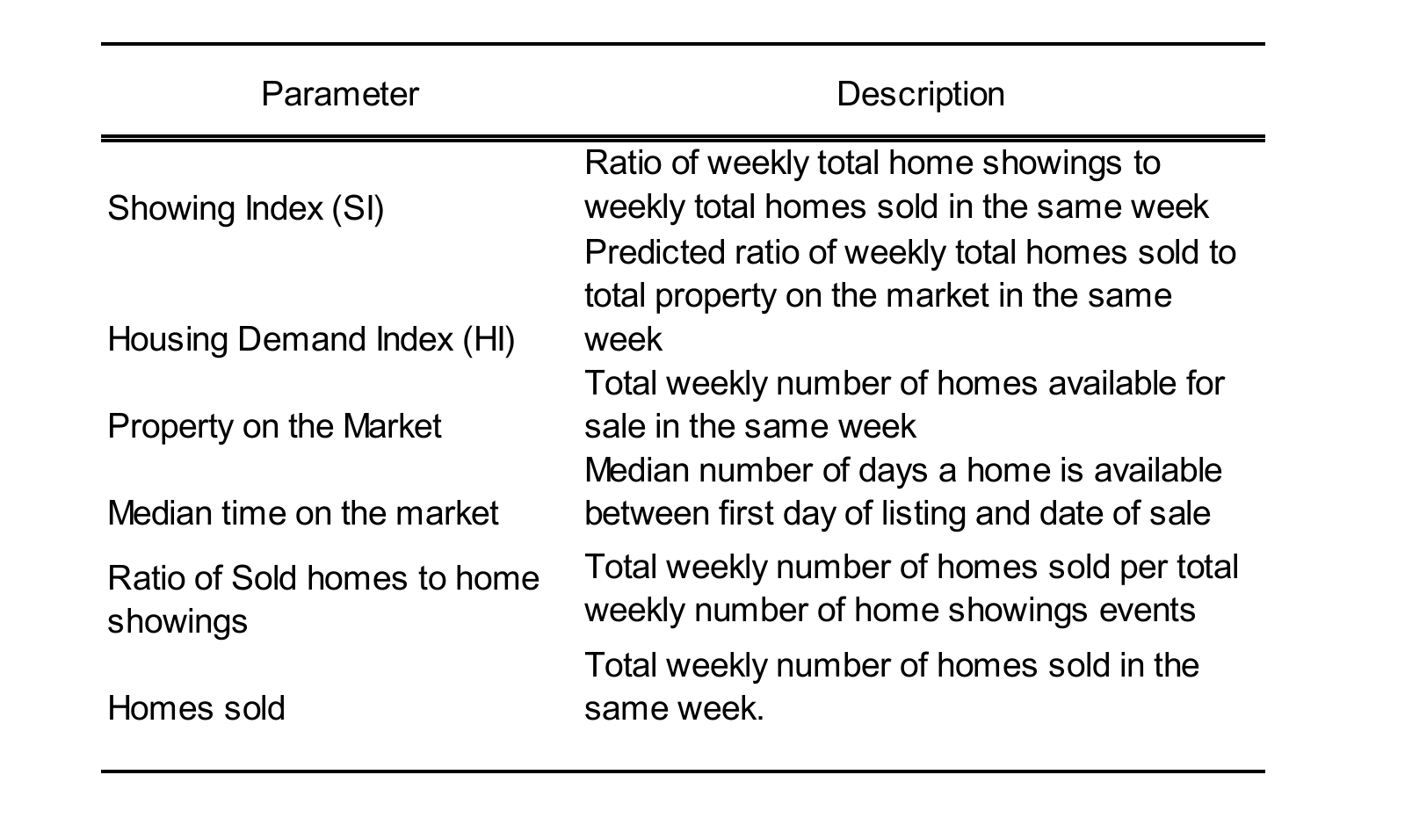}
\caption{Showingtime Data Parameter Description. Each of the parameters listed above were aggregated to measure weekly totals that comprise the parameters of the dataset.  }
\label{tab:parameter} 
\end{table}

\begin{table}[H]
	\centering
	\includegraphics[width=0.8\linewidth]{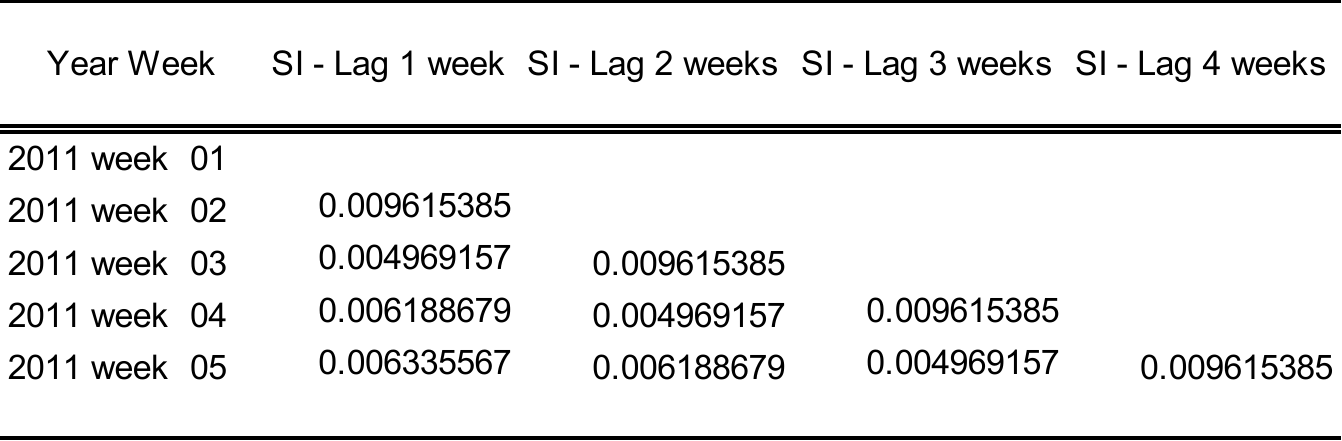}
	\caption{Excerpt From Showingtime Final Data Structure}
	\label{tab:agg}
\end{table}

\begin{table}[H]
	\centering
	\includegraphics[width=0.8\linewidth]{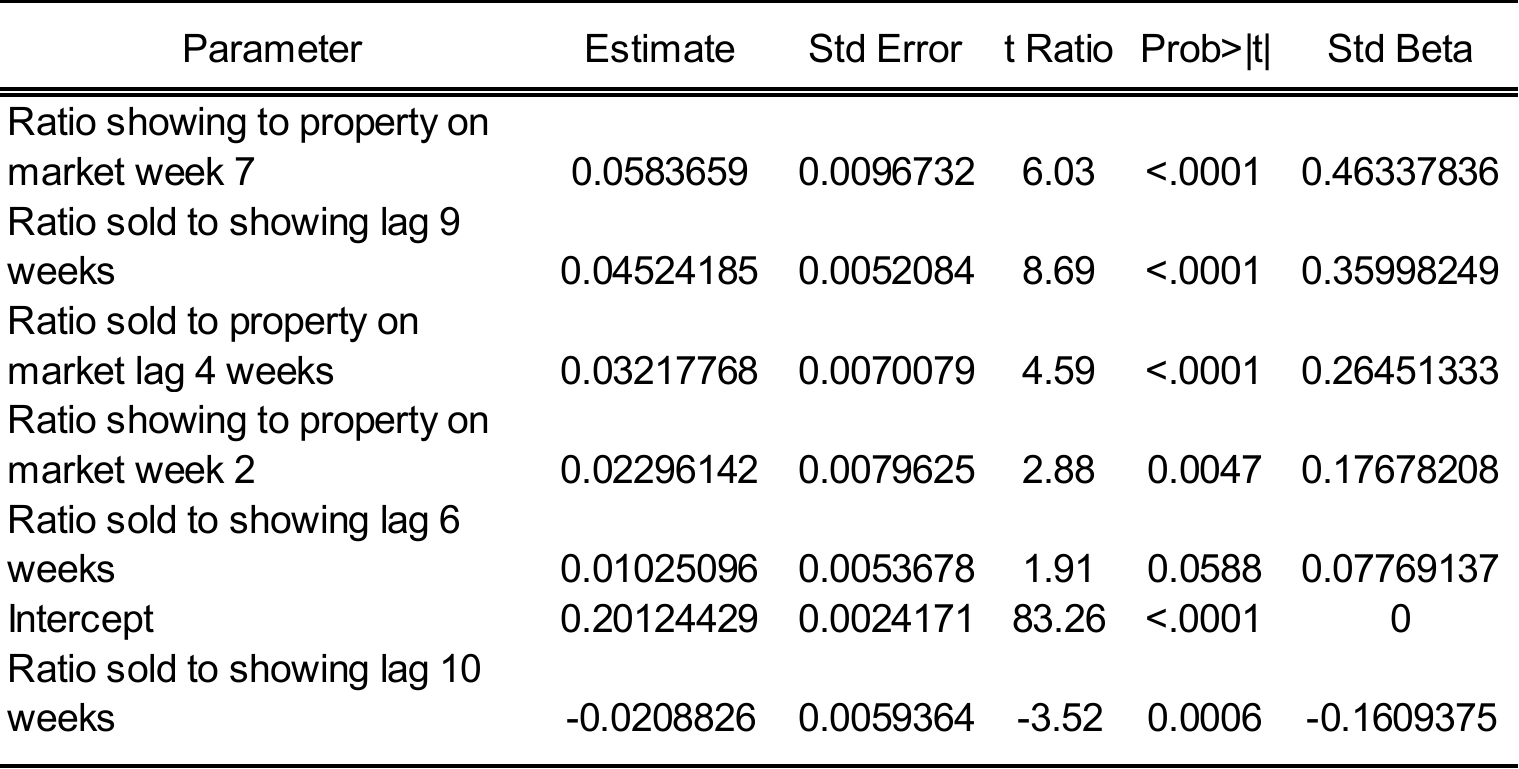}
	\caption{Parameter Estimates From Forward Stepwise Regression Model. Forward stepwise regression was employed to select significant parameters to model the linear relationship between showings and homes sold on a weekly basis in the Chicagoland housing market. }
	\label{tab:forward}
\end{table}

Forward stepwise regression was employed to select significant parameters to model the linear relationship between showings and homes sold on a weekly basis in the Chicagoland housing market. 
We elect to transform the HDI in order to correct for the slight skew in the measure density. Transforming the target parameter has advantages, including a more accurate sample mean, and normal variance and a square root transformation results in a more normally distributed dependent HDI parameter.  
As current week showings and number of home sold depend upon the previous weekly showings - our SI measure, and number of homes sold, we include lagged parameters (as shown in Table~\ref{tab:forward}) from previous week home activity and market activity to account for the impact that previous weeks of demand would have on future demand for a home. These comprise some of the independent variable parameters listed in Table~\ref{tab:parameter}. The full set of predictors include the prior weeks of SI, sales, and property on the market. The outcome of interest is the predicted HDI in subsequent weeks. 
We first split the showings dataset into training and validation datasets with an 80\% -20\% split. We then fit a linear model to the training data in the form of: 
\[
y = \sum_k \beta_k x_k
\]

where y is the HDI, and xk includes lagged SI, lagged HDI, and median time on the market.
To model HDI, forward stepwise linear regression was fitted to the training data set. The fitted model has an adjusted $R$-squared = 0.9. As shown in Table~\ref{tab:forward}, independent variables selected by the forward stepwise algorithm, include several weeks of prior home showings. It also becomes apparent that some previous weeks of showings are more predictive than others; namely, SI weeks 2, 4, 6, 7, 9 and 10 are predictive of future home sales. This is not unexpected given the time needed to negotiate and arrange the final sale of a home. 
We then perform 10-fold cross validation. As a check on this regression model, the model's fit to the training data set was tested on a subset of data not included in the initial model to assess the out-of-sample predictive power. For the training set, 115 weeks of data were randomly selected and used to model the forward stepwise regression. A remaining 29 weeks were reserved as a testing set by which the model could be externally assessed, representing 80\% and 20\% of the total dataset respectively. We again observe that the model captures 90\% of the variance in the training set and 77\% of variance in the testing dataset
\[
\mathrm{HDI}_{EM} = 0.15\times \mbox{Linear Regression} + 0.05 \times \mbox{CART} + 0.8\times\mbox{Neural Net}
\]

\subsection{Regularized Linear Model}
To confirm the importance of the predictors selected by the forward stepwise regression, we fit a    Tikhonov regularized linear regression, also called Lasso \cite{hastie2005elements} to the data set. The model coefficients are found via solving 
\begin{align*}
\mbox{minimize}\quad & ||X\beta - y||_2^2\\
\mbox{subject to} \quad & ||\beta||_1 \le \lambda
\end{align*}
where $\lambda > 0$ is a scalar regularization parameter, $X$ is the design matrix and design $y$ is the vector of observed values of HDI index.  We fitted the model with a total of 35 predictors, including the SI, 5-week to 20-week lagged SI, median time on the market, week, and 5-week to 20-week lagged HDI. Instead of using a a quadratic programming solver to find the coefficients, we used the least angle regression (LAR) algorithm, which exploits the special structure of the lasso problem, and provides an efficient way to compute the solutions simultaneously for all values of $\lambda$ \cite{efron2004least}.  

\begin{figure}
	\centering
	\includegraphics[width=0.8\linewidth]{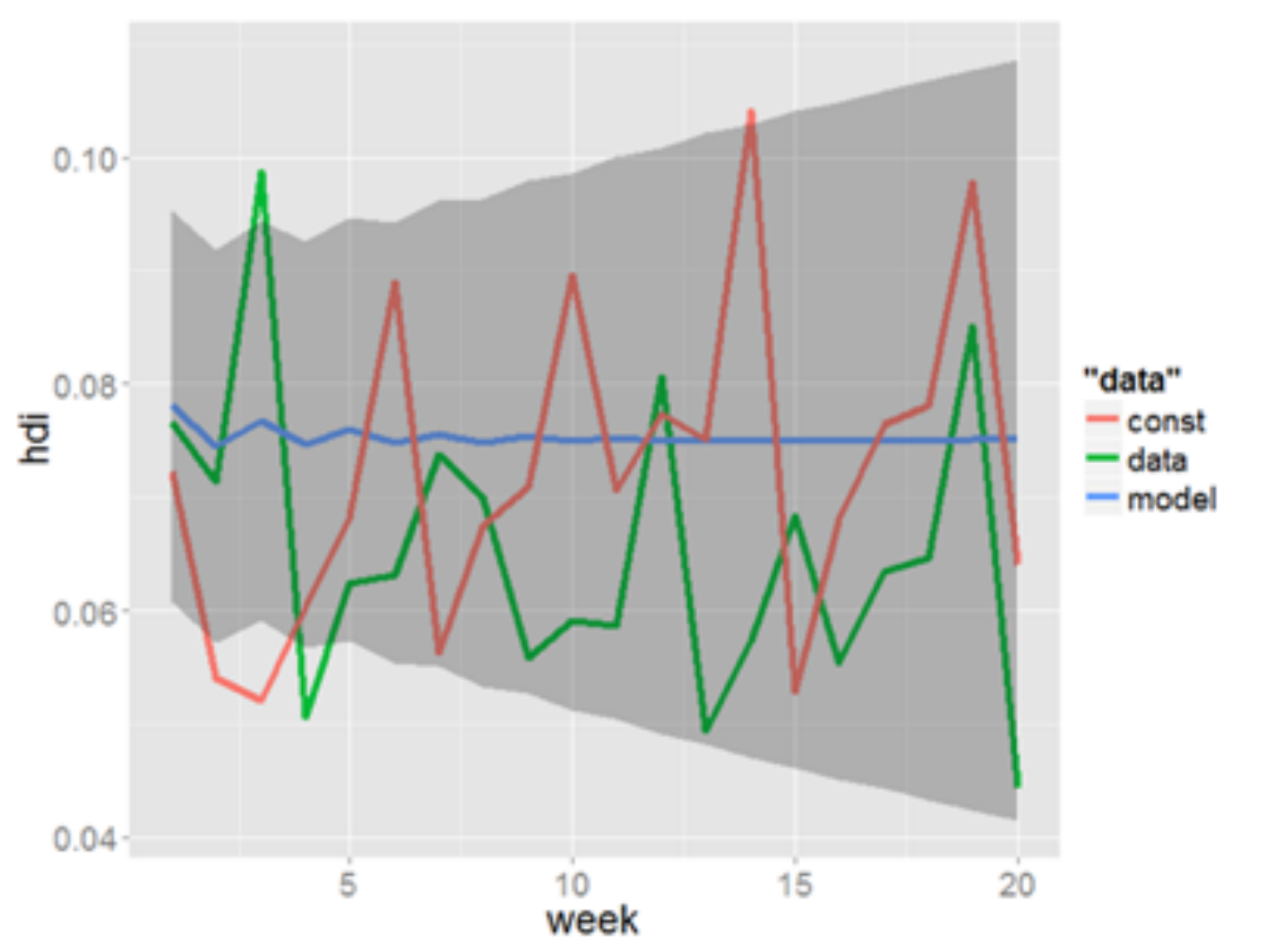}
	\caption{Baseline Univariate HDI Forecast Plot. 20-week forecast using the Univariate time series model. Plot com-pares the data, the forecast from model and naive constant forecast. }
	\label{fig:baseline_forecast} 
\end{figure}

20-week forecast using the Univariate time series model. Plot compares the data, the forecast from model and naïve constant forecast. 
LAR moves the coefficient of the most correlated predictor continuously toward its least-squares value until a second predictor catches up in terms of correlation with the residual. Then the coefficients of both predictors move together in a way that keeps the residual equally correlated with each predictor until the third predictor becomes equally correlated with the residual. The process continues until all the predictors are included the model. Using training data, the most important predictors identified by both Lasso and LAR are SI-L7, HDI-L13, HDI-L8, HDI-L9, SI-L5, and HDI-L17. The MAPEs of test set are 23.57\% for both models.

\subsection{Ensemble Model}
In addition to the linear relationship, we find that statistical learning techniques also capture variance in the models due to non-linear relationships. Decision tree techniques select parameters that resemble those selected in the forward stepwise linear regression model though with some variations. For instance, the first split on week 3's ratio of showing to property on the market is not found to be highly predictive in the forward stepwise regression model. However, the importance of weeks 4, 6, and 9 are also reflected in the decision tree model. 
We find that neural nets in particular are effective in modeling housing demand, predicting 84\% of the validation set variance in home demand, more robust than either linear regression or decision trees. To capture the strengths of the various techniques, we combine forward stepwise regression, decision trees, and neural nets to create the final ensemble model. This ensemble model was developed with the objective of maximizing the adjusted $R$-squared using the testing data set resulting in a final out-of-sample predictor that captures 87\% of the variance in the housing demand.  

\subsection{Treating Time Correlated Errors}
\begin{figure}
	\centering
	\includegraphics[width=0.8\linewidth]{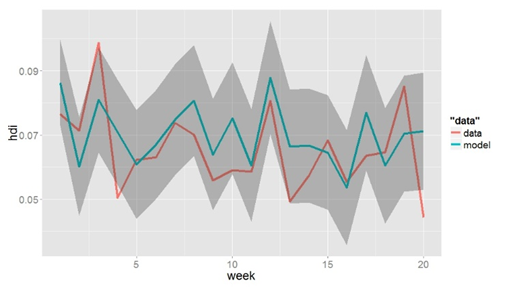}
	\caption{Time Series HDI Forecast Plot With Predictors. 20-week forecast using the time series with exogenous predictor model.}
	\label{fig:forecast-pred} 
\end{figure}

20-week forecast using the time series with exogenous predictor model.
Though models developed in previous subsections can be used to perform an accurate forecast of the housing demand, the model errors do not satisfy a basic assumption of being independently distributed normal variables.  The errors feature serial correlation. Thus, using time series models to model the self-correlated errors provide an additional opportunity to improve the forecast. 
We developed a linear model with correlated errors. As previously, we use linear model for HDI:
Instead of assuming normal distribution of errors, we model those using a time series model. We used a classical seasonal auto-regressive integrated moving averages model (ARIMA) \cite{shumway2013time}. The ARIMA($p,1,q$) model for errors is as follows:

 \[
 z_{k-1} - z_k = \alpha_1z_{k-1} + ... + \alpha_pz_{k-p} + w_k + \phi_1w_{k-1} + ... +\phi_qw_{k-q}
 \]

Our time series data contains an additional yearly periodic component. We assume that demand forecast depends on values at annual lags (i.e. data from the same week from last year, and perhaps two years ago). Thus we include those in the model to obtain a seasonal model with seasonal frequency of 52 weeks, which is an approximation of a yearly cycle. First we test the pure time series model, without any regressors, or where all betas are zero. We fit the univariate time series model HDI, using ARIMA(0,1,3)(0,1,0)[52] and established baseline forecast as shown in Figure~\ref{fig:baseline_forecast}. The ARIMA model was chosen using cross-validation to minimize the Akaike information criterion with correction (AICc). As we can see in Figure~\ref{fig:baseline_forecast}, the model forecast (denoted with a blue line) is a flat line, meaning there is barely any signal in the HDI data itself and it is impossible to forecast housing demand by just looking at historical values. This forecast is as good as using the mean forecast, or forecasting demand in 20 weeks is the same as the average demand over the previous few weeks. Note that the mean forecast (pure ARIMA) outperforms the constant forecast (forecasted demand in 20 weeks is the same as demand in the current week) by 5\%. This shows importance of using other predictors, such as showing data, or SI, that are assets in building a meaningful forecast model.  The 20-week forecast of HDI featured a MAPE of 41.92\%.  
Thus, to improve the forecasting accuracy, we built the ARIMA model including the lagged SI (SI-L5 to SI-L20) as exogenous predictors. While the new model is more complex, it captures the influence of the showing to the demand. The new fitted model (ARIMA(3,1,1)(0,1,0)[52]) with exogenous predictors has superior performance. The MAPE for 20-week forecast lowered to 16.55\%. As shown in Figure~\ref{fig:forecast-pred}, in comparison to the univariate time series model, not only are the confidence intervals of the forecasts tighter, but the forecasts are closer to the actual data.
\subsection{Forecasting Using Fourier Terms}
\begin{figure}[H]
	\centering
	\includegraphics[width=0.8\linewidth]{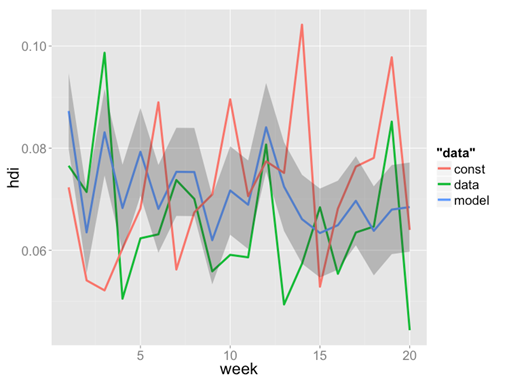}
	\caption{Time Series Plots of Showings by Date. 20-week forecast using the time series with exogenous predictors and Fourier terms. }
	\label{fig:fourier}
\end{figure}

When we forecast the weekly data, Fourier terms are a useful technique for dealing with cyclic data. In the case of weekly observations, the seasonal period is long and noninteger (there are 365.25/7 = 52.18 weeks in a year), so a typical periodic forecasting ARIMA model or exponential smoothing do not appropriately treat such cyclicity even when 52 is used as an approximation of number of observations per year. Further, there can be cycles with different length, such as quarterly or monthly cycles. Fourier terms in the linear forecasting model allow for dealing with non-integer cycle lengths and multiple cycles of different lengths.  The general formula for the model is as follows 
\[
y_t = bt + \beta^Tx + \sum_{j=1}^{K}\left(a_j \sin\left(\dfrac{2\pi j t}{52.18}\right) + b_j \cos\left(\dfrac{2\pi j t}{52.18}\right) \right) + \eta_t
\]
One  of  the  more  straight-forward  approaches  is  to  use  a  dynamic  regression  model  with  Fourier  terms  as predictors  and  modeling  errors  as  an  ARIMA  process.   Results for the ARIMA model with exogenous variable (lagged SI) and Fourier terms is shown in Figure~\ref{fig:fourier}. 

We select the number of Fourier terms using cross-validation and choose the model with the lowest AICc. The order of the ARIMA model for the errors is also selected via minimization of AICc.  The error of the forecast of a model with Fourier terms is comparable with the model with just exogenous variables. Including Fourier terms improves the quality of the forecast, lowering the percent error to 17\% compared to 28\% when a constant forecast is used. One of the main  advantages  of  including   the  Fourier  terms  in  the  model  is  that  it  makes the residuals a stationary process and this allows use of a non-integrated ARIMA model to describe residuals dynamics.  The advantage of having a stationary residuals process is that prediction intervals are more narrow and do not grow as fast as when models with random walk components are used. In fact, the forecast interval is a constant after a few forecast steps \cite{shumway2013time}. In the model selection process, we restrict ourselves to a regression model with ARIMA errors because it has  been shown  that  there  are  some  theoretical  issues  when  using  exponential  smoothing  models  with  the regressors, namely those models are not forecastable \cite{osman2015exponential}. The process is only forecastable if the random disturbance at time t can be expressed as a converging sum of past values. In other words non-forecastable models are mathematical artifacts and do not have any practical meaning. 

\section{Conclusion}\label{sec:conclusion}
We show that a predictive model can be used to forecast housing demand. We demonstrate that both linear and non-linear models can be built that use showing time data as predictors for future demand. The simplicity of the linear models lends itself to an understandable, replicable model for potential use among real estate market decision-makers. We then use time series forecasts to create a longer term forecast of up to 20 weeks by treating the errors as time-correlated as well as testing Fourier terms. Inclusion of Fourier terms into a time series model allows forecasting with  narrower uncertainty intervals. 
Further research into the implications of the HDI is needed to fully understand the possibilities and limitations of home showings as a data source for economic models of the housing market. The value of home showings data is clear in the context of a market heat index model and a housing index model, but home showings might be used to model other relationships in the housing market. Further, they might enhance existing market indicators, or in different market contexts under- or over-represent consumer demand. 
In addition to the usefulness in application to existing indicators, exploring to what degree geographic granularity is possible will also be useful for real estate stakeholders. Where users of existing indices are confined to preexisting geographic boundaries, the home showing data can be matched to a variety of geographic identifiers, including neighborhood, community, or even blocks. The question remains as to how geographically granular and at what loss of predictive accuracy home showings can forecast. Further research should also seek to replicate this work in cities beyond Chicago to ensure the external validity of home showings data in divergent real estate markets. 


\section*{References}

\bibliography{ref}

\end{document}